# Communication with family and friends across the life course

Tamas David-Barrett[a,b,c], Janos Kertesz[d,e,f], Anna Rotkirch[c], Asim Ghosh[e], Kunal Bhattacharya[e], Daniel Monsivais[e], Kimmo Kaski[e]

[a] Kiel Institute for the World Economy, Kiellinie 66, D-24105 Kiel, Germany
[b] Department of Experimental Psychology, University of Oxford, South Parks Rd, Oxford OX1 3UD, UK
[c] Population Research Institute, Väestöliitto, Kalevankatu 16, Helsinki 00101, Finland
[d] Central European University, Center for Network Science, Nador u. 9, Budapest, H-1051, Hungary
[e] Department of Computer Science, Aalto University School of Science, P.O.Box 15500, 00076 Finland
[f] Department of Theoretical Physics, Budapest University of Technology and Economics, H1111, Budapest, Hungary

**Corresponding author:** Tamas David-Barrett, tamas.david-barrett@ifw-kiel.de
Kiel Institute for the World Economy, Kiellinie 66, D-24105 Kiel, Germany

**Abstract**
Each stage of the human life course is characterised by a distinctive pattern of social relations. We study how the intensity and importance of the closest social contacts vary across the life course, using a large database of mobile communication from a European country. We first determine the most likely social relationship type from these mobile phone records by relating the age and gender of the caller and recipient to the frequency, length, and direction of calls. We then show how communication patterns between parents and children, romantic partner, and friends vary across the six main stages of the adult family life course. Young adulthood is dominated by a gradual shift of call activity from parents to close friends, and then to a romantic partner, culminating in the period of early family formation during which the focus is on the romantic partner. During middle adulthood call patterns suggest a high dependence on the parents of the ego, who, presumably often provide alloparental care, while at this stage female same-gender friendship also peaks. During post-reproductive adulthood, individuals and especially women balance close social contacts among three generations. The age of grandparenthood brings the children entering adulthood and family formation into the focus, and is associated with a realignment of close social contacts especially among women, while the old age is dominated by dependence on their children.

**Significance:** We identify the basic types of close social relationships to parents, children, friends, and spouses, from mobile phone communication patterns. The patterns of investment in these close social contacts vary with the six main life stages of adulthood. Women play a more central role in holding different generations of the family together, and are affected more by grandparenthood than men.

**Introduction**

Humans live their lives in stages characterised by distinctive patterns of social relations. Despite sociocultural variation, the basic pattern of life-course dependent sociality is universal. Infants grow to be children, juveniles, young adults ready for reproduction, then the majority pairs up, becomes parents and raises children, many live long enough to become old, and eventually we all die. During these natural phases, humans, like many other animals, have social relationships reflecting their dependence on and investment in family and peers. First with parents and siblings, then increasingly with peers and lovers, typically followed by union formation and transition to parenthood, and later the transition to grandparenthood and old age. During these stages not only do we have different patterns of social relationships around us, but the function and intensity of these relationships change (1), partly reflecting gender differences in reproductive strategies.

While anthropological evidence shows remarkable universality of the main life course stages across different cultures (2, 3), surprisingly few studies have investigated how social ties in contemporary societies evolve across the entire adult life course (4). Here, we study the way the human life cycle is associated with relations to the closest social contacts depending on relationship type, using a large database of mobile communication in one specific European country.

Nowadays much of the interpersonal communication goes over mobile phones, the coverage of which in developed countries is close to 100 % of the adult population. Therefore, call records enable detailed tracking of relations between the closest ties (5-7). The frequency and length of phone calls reflect the strength of the tie between callers in the sense of Granovetter as they are related to the time and financial investment (7-9). Moreover, the party initiating the call can be considered to be more motivated in maintaining the contact than the receiving party (7, 10, 11). Calling patterns can inform us about cross-generational family relations (7) and spatial distribution of close social ties (10). However, since previous studies have not methodologically separated family ties from non-kin ties they have been unable to investigate how various stages of the family life course vary by relationship type.

Here, we distinguish between three dyadic bonds crucial for human sociality: parents and children, romantic partners, and same-sex friends. We investigate how communication within each such tie is associated with six main life stages of adulthood: early adulthood, union formation, middle adulthood, post-reproductive adulthood, grandparenting, and old age. (See Data and Methodology-section.).

We are especially interested in gender differences across different life stages and the effect of grandmothering on social behaviour. Across societies, compared to all other caretakers mothers tend to provide most child care to their infants and young children (12) through a family bond, which is crucial for child outcomes in later life (13, 14). Mothers also tend to remain emotionally closest to their children and especially their daughters as these grow up and have children themselves (e.g., 15); this general preference for maternal kin persists in contemporary Europe (16, 17). In previous research on mobile phone communication patterns, it has been demonstrated that women are more nepotistic in their phone call patterns then men: they call a smaller circle of contacts, but more intensively (2, 18, 19).

The importance of grandmothering in humans suggests that women alter their reproductive strategy dramatically at the age of menopause, whether as the result of a specific evolutionary adaptation (20) or as a by-product from other evolutionary forces (21). While pre-menopausal women focus on producing and raising their own offsprings, post-menopausal women focus on providing alloparental care to their grandchildren, a form of care which has been crucial to human development (22) and remains important for child wellbeing (14, 23). Men, by contrast, do not have a similar clear shift in their reproductive capacities. Their function as grandparents is also different: While grandfathers may also be important for child survival and well-being, the presence of grandfathers has more often been related to no benefits for grandchildren or even to adverse grandchild outcomes (14, 23-25). The behavioural implications of this gender difference in modern societies have not been previously explored (but see 30).

The family life course relies on four close social bonds (26): the parent-child dyad, the sibling relationship, the spousal dyad, and the relationship between friends. Siblings have to be excluded here, for reasons explained in the data section. Friendship is defined as a tie between two individuals who are not relatives, of the same sex, and not romantically involved. In this data, we further defined friends as being of similar age, allowing us to differentiate between friends and siblings, since the latter typically have an age difference of one year or more. This allowed us to identify six roles for the ego in relation to specific alters: 'mother', 'father', 'friend', 'spouse', 'son', and 'daughter'. The parent-child bond may also imply grandparent-grandchild relationships provided that the life span is long enough. Note that in this study a family generation is denoted by the age difference of around 25 years, friends are confined to same-sex alters of same age, and spouses are confined to opposite-sex alters of similar age (see Data and Methodology section).

We assume that the ego's age-dependent life stages defined above will be associated with different communication patterns among different age individuals, and study the following three research questions:
While social network analyses consistently show that peer relations dominate adolescence and young adulthood (27-29), the transition to parenthood, to old age is related to changes in both quantity and quality of social relations (2, 30, 31). We investigate (i) *the differences in relative emphasis in communication with peers (friends and romantic partners), and with kin (parents and their adult children) throughout adulthood.* Second, we hypothesise that women's communication with kin has more prominent role within the kin network compared to that of men (27, 32). For adults this implies different foci from men and women on the members of their close ego network during early adulthood, mature adulthood, and grandparenthood. In particular, due to the importance of mothering and grandmothering (20), we expect the (ii) *cross-generational contacts of female egos to make up a relatively higher proportion than those of the male egos, independent of age*. Third, due to the importance of grandmothering, we hypothesise (iii) *that egos of grandparenting age will have bigger gender differences in their behaviour compared to younger egos, and that females of grandmothering age will exhibit more calls towards their adult children compared to men*.

**Results**

The phone call pattern when clustered into bins by the age of the caller exhibits a generational effect for both men and women (Fig. 1). For instance, for a 30-year-old ego the great majority of the phone calls are conducted with the similar age alter. There is also a second, smaller peak in the distribution, namely to alters who are one generation, i.e. 25 years, older. Among older egos this generational peak also increases with age, so that calls to an alter of the same age is present among egos of different ages. From the age of 45 three distinct generations appear: one for the ego's own generation, one for the older generation, and a third one for a younger generation. Among older egos, the peak to the older adult generation is smaller, while the peak to the younger generation is bigger, compared to younger egos. The pattern is similar both for female and male egos, though more pronounced for females.

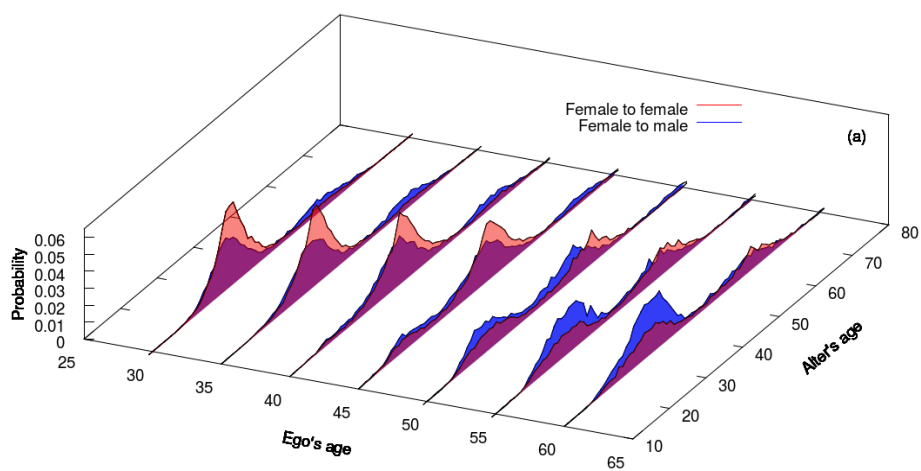

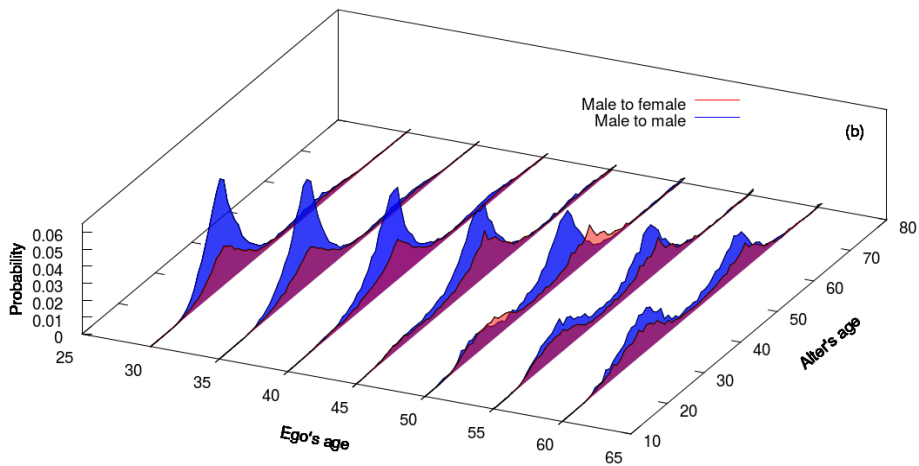

Fig. 1. Age dependent call frequency of most frequently called alters for (a) female and (b) male callers, as functions of the ego's and alter's ages. As the ego's age increases the alter's age also changes, suggesting the presence of three distinctive family generations in the ego's social network.

## Life stage, peers and kin

In Fig. 3 we depict the age dependent phoning patters of egos to their close alters using the number of calls, the average fraction of total phone call time, the balance between out and in calls, and the average length of time per call, as measure. We show that the ego's phoning patterns are in line with their assumed life stage (Fig. 3).

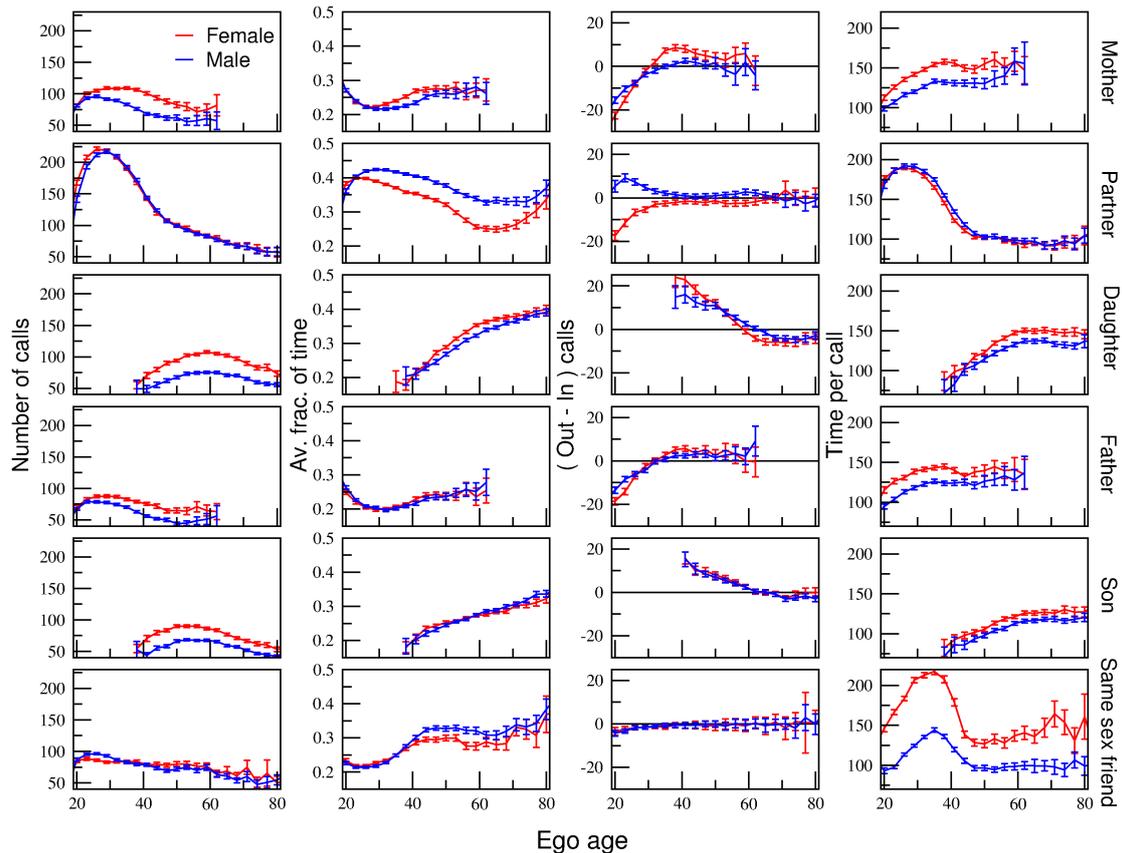

Fig. 3. Age-dependent phone communication patterns of egos with close network neighbors, i.e. "mother", "father", romantic partner or "spouse", "best friend", "daughter", and "son", using four different measures: the number of calls, the average fraction of total phone call time, the balance between out and in calls, and the average length of time per call.

In this study we separate the period of *young adulthood* into two phase, i.e. early adulthood for individuals between 18-21 year olds, which we call early adulthood and between 22-28 year olds which we identify with first union formation. The frequency of phone calls and call length to alters increase for both these age-windows. Thus, the early adulthood looks like a precursor to union formation: the dynamics are the same, but to a lesser extent among the younger. The number of phone calls to all alters and call lengths also increase to all alters. The change in communication pattern for both the call frequency and length is thus characteristic for the entire period of young adulthood, or for individuals between 18 and 28 year olds. However, there is an important difference between these two periods. In the early adulthood phase the average fraction of time spent talking to either the "parents" or the "best friend" is falling, while the fraction of time talking to the "spouse" is increasing. (Note that the "spouse" at this age is likely to be a romantic partner to whom the ego is not married, yet.) In other words, although the ego speaks more frequently and for longer times to parents, friends, and romantic partners, he or she speaks increasingly more to the romantic partner. Calls from ego's own parents are most frequently received in the young adulthood phase.

The second phase of the young adulthood life stage is characterised by finding a long-term romantic partner, and creating a strong romantic bond with him or her. In this, the ego and partner form a family, in which the ego relies especially on the communicational support from the same-sex "best friend" (33).

Following the young adulthood comes the stage of family formation and maintenance during middle adulthood life stage for the 29 to 45 year olds, which is characterised by decreasing communication with the "spouse". In contrast there is increasing communication with the "best friend". Note that although the call frequency peaks for about the 25 year olds, while the fraction of call time increases with old age such that the average length of the call to the friend peaks for about 35 year olds. This difference in peer communication is probably partly explained by the fact that spouses have moved to living together without the need to communicate each others by making phone calls, while there may be less time to meet friends face-to-face.

At this stage there is also a difference in the communication pattern with the ego's parents compared to younger egos. Not only is the average fraction of time talking to the parents bigger, but, crucially, in this period also the direction of initiating the phone calls is the reverse. While the years before the person reaches mid-30s are dominated by the parents overwhelmingly initiating phone calls to the ego. Among the egos in their mid-30s, he or she is more likely to call the parent than vice versa. For the period when the ego is typically having a family with young children, he or she increasingly appears to rely on the support from her parents and best friend in communication patterns.

Following the middle adulthood comes late adulthood and old age, or more generally post-reproductive adulthood, i.e. past the period of having children. We divide this period in the ego's life cycle into three phases. The first of these we define as between the age of 46 and 55, during which the life of the ego is characterised by children who are leaving childhood and juvenility and entering adolescence and young adulthood themselves. At the same time, the parents of the ego are still likely to be alive, and hence in this period the ego is juggling three generations of close contact: his or her spouse and best friend, his or her parents, and his or her children (Fig. 4).

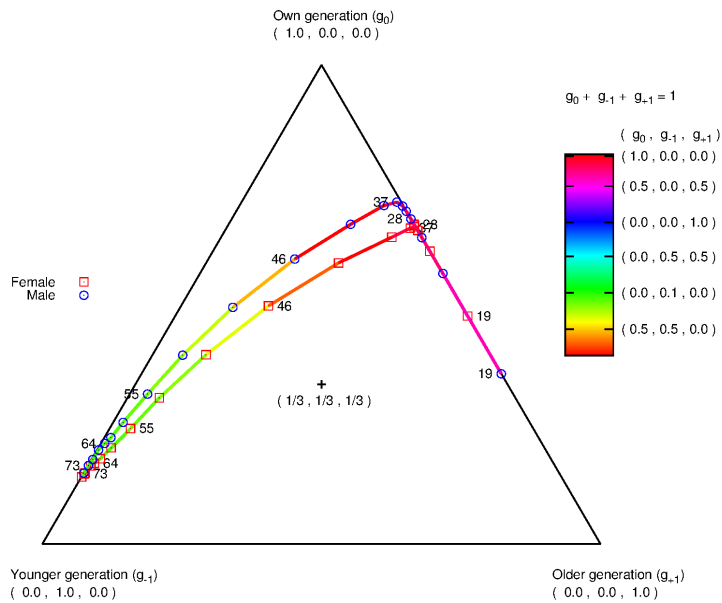

Fig. 4. Intergenerational balance of phone calls. The distance from each point to the centre of the triangle represents the imbalance, measured as $\left((1/3-g_0)^2+(1/3-g_{-1})^2+(1/3-g_{+1})^2\right)^{0.5}$, its location in the triangle shows towards which generation the imbalance is shifted. Here g0, g-1, and g+1 represent the fraction of calls that the egos in a certain age group made with alters in their own, older, and younger generation respectively. Therefore, the imbalance between these three fractions for each age group is mapped to a point in the triangle. The shifting is also represented with a colour map. Small numbers at the symbols indicate the age of the ego.

Next, the second phase of post-reproductive adulthood occurs at the ego's ages of 56-75, which is the period during which the ego is most likely to be a grandparent with his or her grandchildren being in their infancy or childhood or juvenility. The grandparenthood is characterised by a radical realignment of ego's relationship with the alters. One characteristic is that during this period the parents of the ego are starting to pass away thus causing the stop of communication but preceded by an increased call frequency. The second characteristic is that the pattern of communication to the children, who are at this point in their late 20s to late 40s also changes. The ego increases the core frequency to his or her children, especially to his or her "daughter". The increased focus on the children is associated with reduced communication with all other alters. Interestingly, in this age group the call initiation pattern with the children changes. Unlike younger ego networks, egos of grandparental age are more likely to be called by their children than to be initiators of the calls (Fig. 3).

Finally, the third post reproductive adulthood phase, old age, are the years after 75. Egos of this age focus on their own generation. The average fraction of time talking with either the spouse or the best friend is again bigger compared to younger egos. This is interesting, since living arrangements have not changed much compared to the previous life stage. The decreased balance between in- and outgoing calls evident at earlier life stages (in which the children were increasingly more likely to initiate a phone call) is more even among ego networks of this age compared to younger age groups.

**Gender differences in life stage patterns**

Apart from the variation of communication patterns with life-course stages, we also observe important gender differences in the apparent role of the egos. First, from the middle of the young adulthood phase, female egos are more likely to have cross-generational communication than male egos. This was the case also after controlling for the fact that women (at least past the age of 26) spend more time talking on the phone in general (Fig. 5, Supplementary Material Fig. S1). This supports our second research hypothesis, that women play a more central role in holding together the different generations of the family.

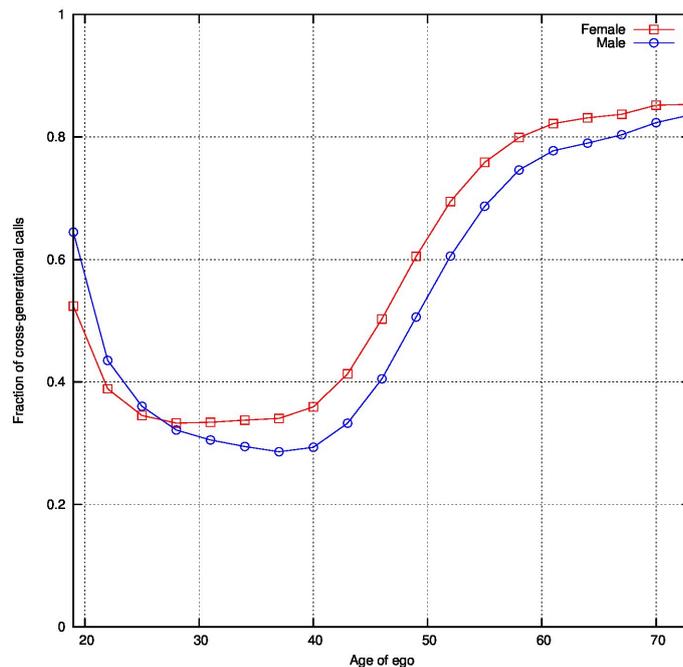

Fig. 5. Gender difference in cross-generational phone calls: the fraction of time of total phone calls that the ego spends communicating with the alters "mother", "father", "daughter", and "son". X-axis: age of ego, y-axis: the percentage of cross-generational phone calls. Red: female, blue: male.

Second, the reallocation of relationship with alters is more pronounced for women than for men during the late adulthood life stage associated with the grandparental role (Fig. 3). While there is little or only small difference in the change of relationship between mothers vs. fathers towards their son during grandparenthood, the relationship with the daughters is affected significantly more in the case of the mother compared to the father. In this period daughters are more likely to initiate a phone-call towards their mother than their father, which is a disproportionate change compared to the previous life-course stages, while the average length of the phone-call similarly increases more between grandparenthood aged females and their daughters compared to males and daughters. This supported our third research hypothesis about the gender differences in grandparenting.

**Discussion**

In this paper we have demonstrated that it is possible to identify some of the social rules of the average mobile phone user's contacts play in the user's life. In particular, it is possible to identify the parents, the children, the spouse, and the best friend of the average ego. Using this methodology, we were able to confirm several hypotheses that are already present in the literature: people go through distinct phases in their lives, all of which have different social relationship and communication patterns; and all people, especially women, have a tendency to rearrange their social lives when they become grandparents.

Our results are limited by several caveats. First, the data is from one calendar year and cross-sectional. Thus differences between age groups may represent cohort differences in social behaviour, not necessarily life stage differences. However, there is no reason to assume that the core of social behaviour has changed so radically during the last decades in the country that e.g. the cultural codes of behaviour among 40-year olds would be very different from that among 30 or 50 year olds. Fortunately for our purposes, mobile calls were the main way of keeping everyday contact during the study year (2007), and had not yet been massively replaced by other contact platforms (e.g. Facebook chat, WhatsApp or SnapChat) that are more generation-bound and not recorded through mobile phone operators.

Nevertheless, mobile phone communication serves as one form of communication among many (34). For example, we expect that life-course changes affect many other channels of communication, naturally influencing also the mobile phone communication pattern. For instance, cohabitation provides an in-person communication channel potentially supressing the mobile communication channel. Similarly, retired couples can be assumed to be more in personal contact throughout the day, resulting in a falling between-spouse phone use. Furthermore and arguably there can be a small variation in the propensity to use other forms of mobile communication, e.g., texts (7), which may to some extent prove to be substituting phone calls.

Second, although the underlying biological dynamics behind the life phase approach suggests universality, our current database comes from one particular year and from one particular population. Hence, we do not intend to claim that our evidence is universal. Nevertheless, we suggest that the life-stage dependent variation of social tie patterns, and in particular the social focus, should be universal for humans and for other social species. Moreover, we emphasize that communication records as available in Big Data are particularly suitable to investigate such phenomena.

**Data and methodology**

We analysed the mobile phone dataset of a whole calendar year in 2007 from a single mobile service provider in a specific European country. The dataset includes more than 3 billion calls. The record of each call contains the time, the duration, and the codes of the ego and of the alter (the other individual involved in the ego's call).

We reduced the dataset in two ways. First, the metadata (age and gender and the type of contract) is available only for a fraction of those egos that are users of the service provider who collected the data. As our methodology is dependent on the information about the age and the gender of both callers, we excluded all calls where this metadata was not available for both parties. Second, there are two types of contact with the data provider: individual contract, and family contract. For the latter, the dataset includes the metadata for only one member of the family. As our methodology requires the presence of the age and gender for each caller, we also excluded those users that had a family contract with the service provider. These two steps filtered out all callers that are not associated with gender and age data, leaving 2.5 million male and 1.8 million female egos in the dataset. As the frequency of the ratio between the number of male and female egos vary with age (see Supplementary Material, Fig. S2), we controlled for this factor.

Most ego-centric social networks in the data contained a specific set of contacts, varying with age: two older network members of different sexes, an opposite-sex peer, a same-sex peer, and 1-3 clearly younger network members (35). Although all mobile communication pattern is inherently noisy, given the specifications below we claim that it is plausible to name these contacts the ego's parents, spouse, best friend and children. This approach builds on the recent findings proving that it is possible to identify some average social relationship patterns from digital communication data (7, 36).

We measure call frequency, call initiation, length of calls, and relative fraction of time spent talking to the particular alter. We assume that the call length and frequency indicate emotional closeness (9) and that the call initiation indicates greater interest in the alter, i.e. signalling emotional or financial need (10). We assume that an age gap of around 25 years between callers represents a family generation.

We approach the life course as a series of stages characterising the existence of a specific population (37). At each life stage, a set of relatively few intimate relationships to family members and friends constitute the main core of the social life of an individual (11). To examine the way the life stages affect close relationship patterns within the ego network, we assume six main phases from young adulthood to old age (4, 26). We assign specific assumed average ages to each phase, based on female averages from the study population. (Here we avoid age overlapping between life phases although they obviously exist in reality.) It should be noted that male averages tend to be 2-3 years later for union and fertility events.

***Young adulthood:*** (i) *Early adulthood* is here considered to range from the age of sexual maturity and contribution to the family economy, and – in the modern western society from which our data originates – start of secondary or tertiary education as well as entry into the labour market. This stage is characterised by high importance of peer networks and "best friends", entry to the "mating market" and first dating experiences. This ranges in age from 18 to 21. (ii) *Union formation* is considered to be the period for being on the "mating market" and searching for a long term partner, the majority of individuals finding a partner and forming a strong romantic attachment which becomes more important than the "best friend" (33), ending with cohabitation with or without formal marriage. The age range is 22-28.

***Middle adulthood*** (iii) is considered to start from the arrival of the first child and ranging to the age of the parent when the last child reaches adolescence and own parents reaching old age. The age range is 29-45.

***Late adulthood*** consists of the initial period of (iv) *post-reproductive adulthood* phase ranging from the children reaching young adulthood to the children finding their own long-term partners. There is menopause for females and own parents reach very old age. The age range is 46-55. This is followed by (v) *grandparenting* signified by the arrival of the first grandchild and their own exit from the labour market and their own parents starting to pass away. The age range is 56-75.

***Old age***  (vi) when most grandchildren leave childhood and onset of old age illnesses. The age range is from 76 till death.

Second, we observed a distinct gender pattern interacting with ages of the egos (Fig. 2).
  i. the peak of communication of the ego with alters of the same sex and same generation
  ii. communication with alters who are of the same generation and different sex peaks with an age difference for both male and female egos, but so that the male age peak is about two years later than the female age peak;
  iii. similarly, there is a difference between the female and male alters' peaks one generation up: the male age peak is two years later than the female age peak;
  iv. there is no difference between the age of the female and male alters' peaks one generation down.

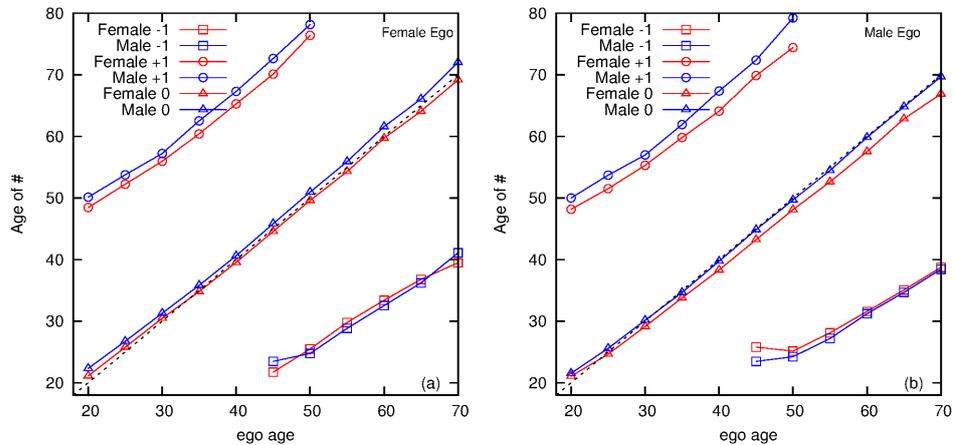

Fig. 2. The age difference between the ego and alters. X-axis: age of the ego, y-axes: age of alter at frequency peak in same generation, one generation up, and one generation down.

These data characteristics indicate that
i. the primary same-sex peer alter is a friend or sibling;
ii. the primary opposite-sex peer alter is a long-term romantic partner, since the age difference between the ego and this alter is of the same size and in the same direction as the spousal age difference within marriages in this specific country as reconstructed from census data;
iii. similarly, the age difference between the peaks of one generation older alters suggests that they are themselves a married couple, which is consistent with the assumption that these will mostly be the parents of the ego.
iv. the younger generation represent egos' children, supported by the fact that there is no age difference between the frequency peaks as there would be between siblings

In principle, these calls may also include non-kin such as friends, neighbours or co-workers, or less related kin such as aunts and uncles or nieces and nephews. Based on previous studies (38, 39) one can nevertheless safely deduce that parental relations will constitute the vast majority of these cross-generational phone calls, and spousal and friend relations will constitute the majority of peer calls.

These observations allow us to distinguish alters that might play particular roles for the ego, assuming that for the average person in this population the close ego network alters are being formed by the close kin, that is, parents, siblings, offsprings, and the closest non-kin friends (30, 40). This is in line with some similar studies in the literature (19).

"Mother": We assume that for the average person the most frequently called alter one generation older is the mother. Hence, we define "mother" as the most frequently called female alter among all alters with ages of 20-40 years older than the ego (41).

"Father": Similarly, we assume that the most frequently called male alter one generation older is usually the ego's father. We define "father" as the most frequently called male alter among those who are 22-42 years older than the ego.

We thus exclude parent-child age differences outside of this age window. This is necessary in order to avoid that a sibling or friend with whom there is a large age difference would be taken into account as a "parent", or that a grandparent with whom there is a short age difference is counted as "parent".

"Spouse": We assume that the most frequently called opposite-sex peer alter is most likely to be the romantic partner of the ego. Thus we define the "spouse" as the same generation alter with an age difference of -2 to 5.

This definition of a romantic partner is problematic as it assumes that there are no homosexual couples in the database, and that opposite-sex peers are likely to be romantic partners rather than either friends or siblings. However, calls to both of these other alter types (homosexual spouses or opposite-sex friends and siblings) can safely be estimated to be much less frequent than calls to heterosexual spouses. Homosexual couples amount to a few percentages in this population, depending on birth cohort (e.g. (42)). Individuals are also much more likely to call their spouses than their siblings or friends (e.g. (43, 44)). Of the cohorts in our study population, over 75 per cent have married by the age of 35 (45).

"Best friend": We assume that independent of age the most likely opposite-sex peer alter of the ego is either going to be a same-sex sibling, or a best friend. Best friends are most likely to be of the same sex and age (46). To separate siblings from friends, we assumed a very narrow age range around the ego's age: only one year. Although this definition does not exclude twins, or same-sex siblings born within the same year, both of these cases are rare. Furthermore, the above definition obviously excludes best friends who are more than a year apart from the ego, which is a price we have to pay for delineating siblings from friends.

"Daughter": Similarly to the way we defined the "parents" one generation older than the ego, we can also identify the "children". Thus we assume that the most frequently called female alter one generation younger than the ego is the "daughter". Just as with the older generation, we define the younger generation as 30 years younger with a +/-10 year window.

"Son": Similarly, we define the "son, as the most frequently called male alter one generation younger than the ego.

Using these definitions, we identified the above alters in the data as the most frequent call partners within the particular gender and age category. When the first ranked alter did not contain any demographic information (for instance due to being with a different phone company), then we took, as a proxy, the next highest ranked alter into the category instead.

**Supplementary Material for
Communication with family and friends across the life course**

Tamas David-Barrett, Janos Kertesz, Anna Rotkirch, Asim Ggosh, Kunal Bhattarcharya, Daniel Monsivais, Kimmo Kaski

**Phone calls with three generations**

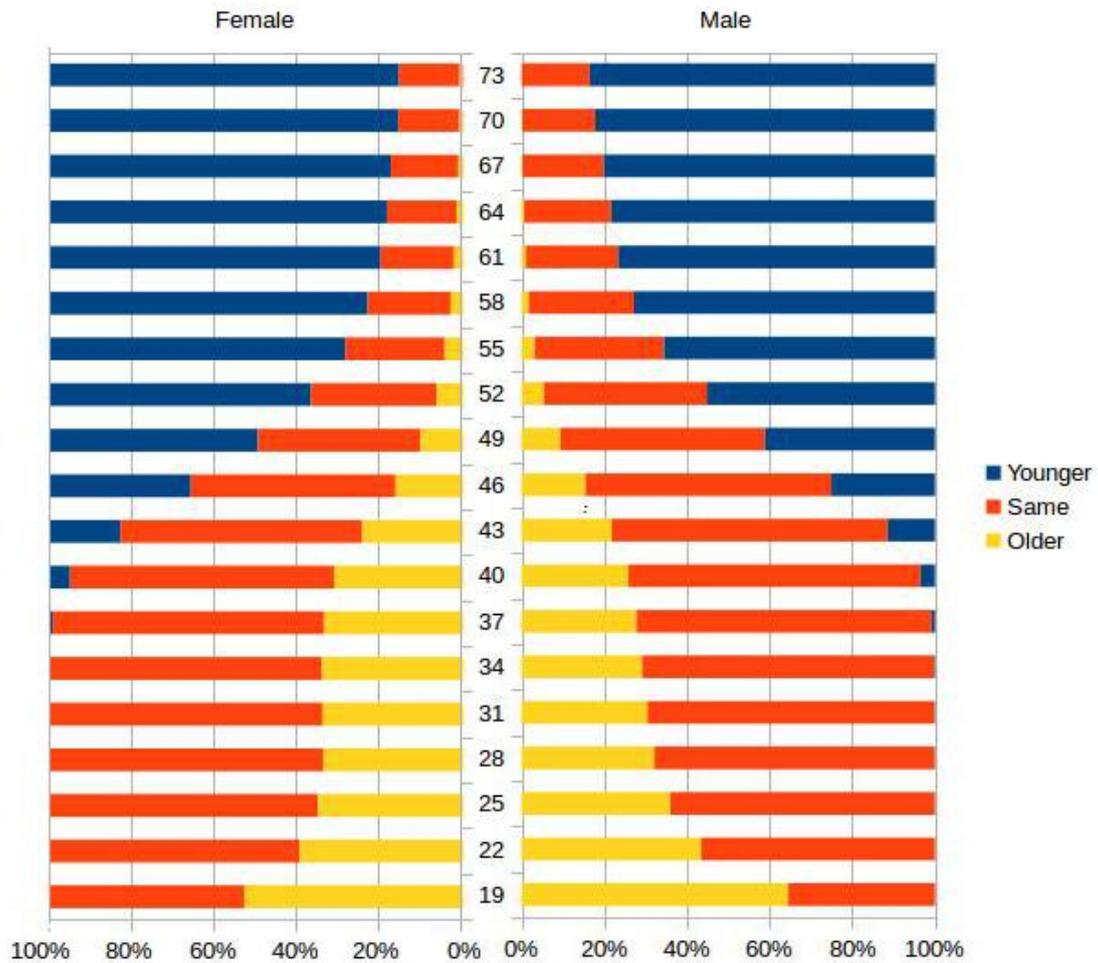

Fig. S1. Histogram of the fraction of the calls with the three generation by female egos (left panel), and by male egos (right panel), as a function of their age (middle panel). Blue, red, and yellow represent the generation of the alters, respectively: one generation younger alter, same generation alter, one generation older alter.

## Gender imbalance in the data

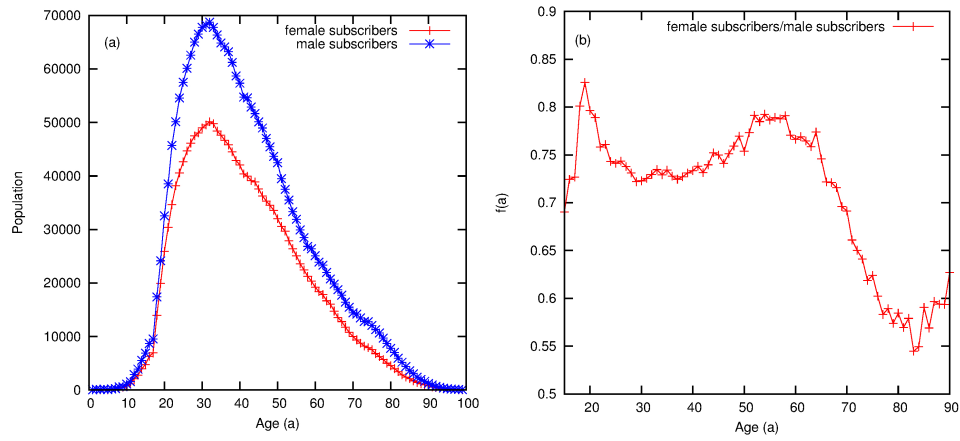

Fig. S2. Frequency difference between female and male callers in the database. Panel (a): histograms, panel (b): ratio between the two genders as a function of age.